\documentclass[aps,prl,twocolumn,showpacs,preprintnumbers,amsmath,superscriptaddress,amssymb,floats,nofootinbib]{revtex4}
\usepackage{graphicx,color}
\begin{document}

\def\Journal#1#2#3#4{{#1} {\bf #2}, #3 (#4)}
\def\NCA{\rm Nuovo Cimento}
\def\NPA{{\rm Nucl. Phys.} A}
\def\NIM{\rm Nucl. Instrum. Methods}
\def\NIMA{{\rm Nucl. Instrum. Methods} A}
\def\NPB{{\rm Nucl. Phys.} B}
\def\PLB{{\rm Phys. Lett.}  B}
\def\PRL{\rm Phys. Rev. Lett.}
\def\PRD{{\rm Phys. Rev.} D}
\def\PRC{{\rm Phys. Rev.} C}
\def\ZPC{{\rm Z. Phys.} C}
\def\JPG{{\rm J. Phys.} G}



\title{ Mass Hierarchy Resolution in Reactor Anti-neutrino Experiments:\\
Parameter Degeneracies and Detector Energy Response}
\date{\today}

\author{X. Qian}\email[Corresponding author: ]{xqian@caltech.edu}
\affiliation{Kellogg Radiation Laboratory, California Institute of Technology, Pasadena, CA}
\author{D. A. Dwyer}
\affiliation{Kellogg Radiation Laboratory, California Institute of Technology, Pasadena, CA}
\author{R. D. McKeown}
\affiliation{Thomas Jefferson National Accelerator Facility, Newport News, VA}
\affiliation{College of William and Mary, Williamsburg, VA}
\author{P. Vogel}
\affiliation{Kellogg Radiation Laboratory, California Institute of Technology, Pasadena, CA}
\author{W. Wang}
\affiliation{College of William and Mary, Williamsburg, VA}
\author{C. Zhang}
\affiliation{Brookhaven National Laboratory, Upton, NY}

\begin{abstract}
Determination of the neutrino mass hierarchy using a reactor neutrino
experiment at $\sim$60 km is analyzed.
Such a measurement is challenging due to the finite
detector resolution, the absolute energy scale calibration, as well as the
degeneracies caused by current experimental
uncertainty of $|\Delta m^2_{32}|$. The standard
$\chi^2$  method is compared with a proposed Fourier
transformation method. In addition, we show that for such a measurement to succeed,
one must understand the non-linearity of the detector energy scale at the level of a few tenths of percent.
\end{abstract}

\maketitle
\thispagestyle{plain}


\section{Introduction and degeneracy caused by the uncertainty in $\Delta m^2_{atm}$}\label{sec:introduction}
Reactor neutrino experiments play an extremely important role in understanding the
phenomenon of neutrino oscillation and the measurements of neutrino mixing parameters~\cite{mckeown}.
The KamLAND experiment~\cite{kamland} was the first to observe the disappearance of reactor
anti-neutrinos. That measurement mostly constrains solar neutrino mixing $\Delta m^2_{21}$ and
$\theta_{12}$. Recently, the Daya Bay experiment~\cite{dayabay} established
a non-zero value of $\theta_{13}$. $\sin^2 2\theta_{13}$ is determined to be 0.092 $\pm$ 0.016 (stat)
$\pm$ 0.005 (sys). The large value of $\sin^2 2\theta_{13}$ is now important input to the design
of next-generation neutrino oscillation experiments~\cite{LBNE,LENA} aimed toward determining
the mass hierarchy (MH) and CP phase.

It has been proposed~\cite{petcov1,petcov} that an intermediate L$\sim$20-30 km baseline experiment
at reactor facilities has the potential to determine the MH. Authors of Ref.~\cite{learned} and Ref.~\cite{zhan,zhan1}
studied a Fourier transformation (FT) technique to determine the MH with a reactor experiment
with a baseline of 50-60 km. Experimental considerations were discussed in detail in
Ref.~\cite{zhan1}. On the other hand, it has also been pointed out that current
experimental uncertainties in $|\Delta m^2_{32}|$ may lead
to a reduction of sensitivity in determining the MH~\cite{andre,deg,parke}.
Encouraged by the recent discovery of large non-zero $\theta_{13}$, we revisit the feasibility
of intermediate baseline reactor experiment, and identify some additional challenges.

The disappearance probability of electron anti-neutrino in a three-flavor model is:
\begin{widetext}
\begin{eqnarray}\label{eq:osci}
P(\bar{\nu_e}\rightarrow \bar{\nu_e}) &=& 1- \sin^2 2\theta_{13} (\cos^2\theta_{12}\sin^2 \Delta_{31}+\sin^2\theta_{12}\sin^2{\Delta_{32}})-\cos^4\theta_{13}\sin^2 2\theta_{12} \sin^2 \Delta_{21} \nonumber\\
&=& 1- 2s_{13}^2c_{13}^2 - 4c_{13}^4 s_{12}^2c_{12}^2 \sin^{2} \Delta_{21} + 2 s_{13}^2 c_{13}^{2} \sqrt{1-4s_{12}^2c_{12}^2\sin^2\Delta_{21}}\cos(2\Delta_{32}\pm\phi)
\end{eqnarray}
\end{widetext}
where $\Delta_{ij} \equiv |\Delta_{ij}|= 1.27 |\Delta m^2_{ij}| \frac{L(m)}{E(MeV)}$,
and
\begin{eqnarray}\label{eq:phi}
\sin\phi = \frac{c_{12}^2 \sin2\Delta_{21}}{\sqrt{1-4s_{12}^2c_{12}^2\sin^2\Delta_{21}}} \nonumber\\
\cos\phi = \frac{c_{12}^2\cos2\Delta_{21}+s_{12}^2}{\sqrt{1-4s_{12}^2c_{12}^2\sin^2\Delta_{21}}}.
\end{eqnarray}
In the second line of Eq.~\eqref{eq:osci}, we rewrite the formula using the following
notations: $s_{ij} = \sin \theta_{ij}$, $c_{ij} = \cos \theta_{ij}$, and using
$\Delta_{31}=\Delta_{32}+\Delta_{21}$ for normal mass hierarchy (NH),
$\Delta_{31}=\Delta_{32}-\Delta_{21}$ for inverted mass hierarchy (IH), respectively.
Therefore, the effect of MH vanishes at the maximum of the solar oscillation ($\Delta_{21}=\pi/2$~\footnote{This is true for $\Delta_{21} = n\pi/2$, with $n$ being an integer.}), and
will be large at about $\Delta_{21}=\pi/4$. Furthermore, we can define
$\Delta m_{\phi}^{2} (L,E) = \frac{\phi}{1.27} \cdot \frac{E}{L}$
as the effective mass-squared difference, whose value depends on the choice of neutrino energy $E$ and baseline $L$.
Since $|\Delta m^2_{32}|$ is only known with some uncertainties
($|\Delta m^2_{32}| = (2.43 \pm 0.13) \times 10^{-3} eV^2$~\cite{PDG} or more recently
$|\Delta m^2| = 2.32 ^{+0.12}_{-0.08} \times 10^{-3} eV^2$~\cite{MINOS}),
there exists a degeneracy between
the phase $2\Delta_{32}+\phi$ in Eq.~\eqref{eq:osci} corresponding to the NH
and the phase $ 2\Delta'_{32}-\phi$ corresponding to the IH when a different   $|\Delta m^2_{32}|$
(but within the experimental uncertainty) is used, namely
$\Delta'_{32} = \Delta_{32} + \phi$ at fixed $L/E$~\footnote{Other degenerate solutions, naturally, might exist when the uncertainty in $ \Delta_{32}$ is larger than $2\pi$.}.
In particular, $\Delta m_{\phi}^{2} (60~km,4~MeV) \approx  0.12\times 10^{-3} eV^2$
(using the experimental values of $\Delta m^2_{21}$ and $\theta_{12}$ \cite{PDG}), which is similar to the size of the experimental uncertainty of $|\Delta m^2_{32}|$. Thus, at fixed $L/E$ determination
of mass hierarchy is not possible without improved prior knowledge of $|\Delta m^2_{32}|$.

\begin{figure}[]
\centering
\includegraphics[width=90mm]{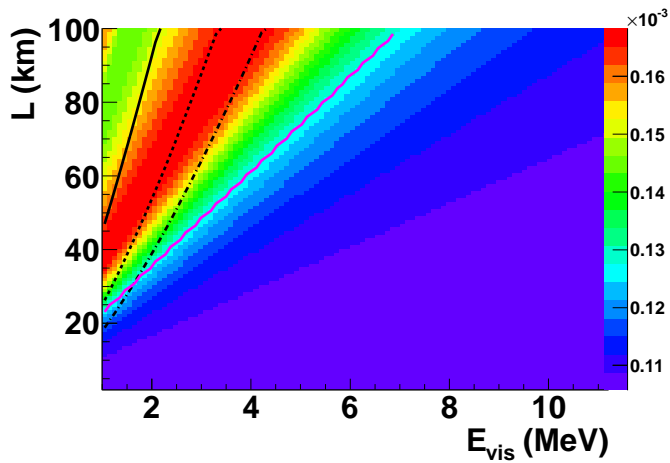}
\caption{ Map of $\Delta m^2_{\phi}$ over a phase space of energy and distance.
The x-axis is the visible energy of the IBD in MeV. The y-axis is the
distance between the reactor and detector. The legend of color code is shown on the right bar,
which represents the size of $\Delta m_{\phi}^2$ in $eV^{2}$.
The solid, dashed, and dotted lines represent three choices of detector
energy resolution with a=2.6, 4.9, and 6.9, respectively. 
The purple solid line represents the approximate boundary of
degenerate mass-squared difference. See text for more explanations.
}
\label{fig:phase}
\end{figure}
\begin{figure*}
\centering
\includegraphics[width=180mm]{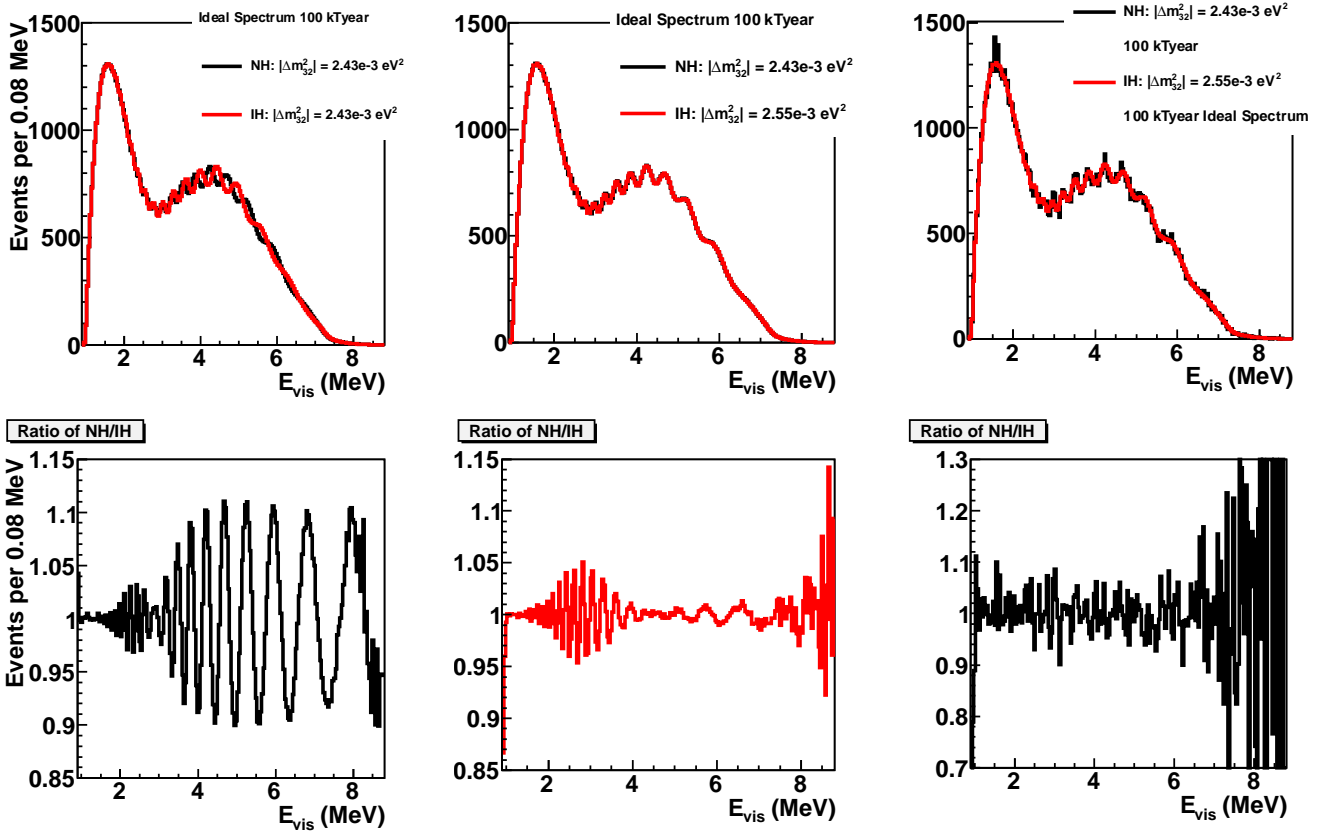}
\caption{\textcolor{black}{Top panels show the comparison of IBD energy spectrum (no statistical fluctuations)
w.r.t. $E_{vis}$ in (MeV) for fixed $|\Delta m_{32}^2|=2.43\times10^{-3}$ eV$^2$ (ideal spectrum
in top left), for degenerate $|\Delta m_{32}^2|$ (ideal spectrum in top middle),
and degenerate $|\Delta m_{32}^2|$ with $100$ $kT\cdot year$ exposure (realistic spectrum
in NH case and ideal spectrum in IH case in top right). The ideal spectrum represents the
case without any statistical fluctuations, while realistic spectrum include these
statistical fluctuations. The resolution parameter $a$ is chosen to be 2.6.
Bottom panels show the ratio of NH to IH case. Due to
statistical fluctuations, the range of Y axis in bottom right panel is enlarged to 0.7-1.3
from 0.85-1.15.} }
\label{fig:spectrum}
\end{figure*}

To some extent, this degeneracy can be overcome by using a range of $L/E$,
or actually, as is the case for the reactor neutrinos, a range of neutrino energies $E_{\bar{\nu}}$.
Fig.~\ref{fig:phase} shows the magnitude
of $\Delta m_{\phi}^{2}$ as a function of distance
between reactor and detector ($L$ in km) and the visible energy of the
prompt events of inverse beta decay (IBD), which is related to the incident neutrino energy
($E_{vis} \approx E_{\bar{\nu}} - 0.8$ in MeV).
It is seen that for the region with baseline $L$ below 20 km, the
effective mas-squared difference $\Delta m^2_{\phi}$ remains almost
constant for the entire IBD energy range. That indicates an irresolvable degeneracy
across the entire spectrum of IBD given the current experimental uncertainty of $|\Delta m_{32}^2|$.
At larger distances, $\approx$ 60 km,  $\Delta m_{\phi}^{2}$ exhibits some dependence
on energy, indicating that the degeneracy could be possibly overcome, as discussed further below.

With a finite detector resolution, the high frequency oscillatory behavior of the positron spectrum,
whose phase contains the MH information,
will be smeared out, particularly at lower energies. For example, at 60 km and 4 MeV,
$2\Delta_{32} \approx 30\pi$ for $|\Delta m^2_{32}|=2.43\times 10^{-3} eV^2$.
Therefore, a small variation of neutrino energy would
lead to a large change of $2\Delta_{32}$.

We modeled the energy resolution as:
\begin{equation}\label{eq:res}
\frac{\delta E}{E} = \sqrt{ (\frac{a}{\sqrt{E~(MeV)}})^2 + 1}\%,
\end{equation}
\textcolor{black}{with choices of a = 2.6, 4.9, and 6.9. The values of 4.9\%, and 6.9\% are chosen
to mimic achieved energy resolutions of current state-of-art neutrino detectors 
Borexino~\cite{Borexino_nim} (5-6\%)
and KamLAND~\cite{chao_thesis} ($\sim$ 7\%), respectively. The value of 2.6\% corresponds to
an estimated performance for an ideal 100\% photon coverage. } \textcolor{black}{In reality, 
an R\&D plan to reach the desired detector energy resolution 
(better than 3\% at 1MeV) has been proposed~\cite{yfwang}.}
Our simulation suggests that the lines defined by the relations
$2 \Delta_{32}  \frac{\delta E}{E} =  0.68\times2\pi$
represent boundaries of the region where the
high frequency oscillatory behavior of the positron spectrum is  completely suppressed.
The solid, dashed, and dotted-dashed lines in Fig.~\ref{fig:phase} show
these boundaries for a = 2.6, 4.9, and 6.9, respectively.
The left side of these lines (lower values of $E_{vis}$) will yield negligible contributions
to the differentiation of MH.

As pointed out above, when $\Delta m^2_{\phi}$ becomes essentially independent
of $E_{vis}$, the degeneracy related to the $|\Delta m_{32}^2|$ uncertainty makes
determination of MH impossible. Again, our simulation suggests that the dividing
line is $\Delta m^2_{\phi}=0.128\times10^{-3} eV^2$, indicated by
the purple line in Fig.~\ref{fig:phase}.
The right side of this line (larger values of $E_{vis}$) {\it alone} will
play very small role in differentiating between these two degenerate solutions.
Thus, the region between the steep lines related to the energy resolution and the
purple diagonal line related to the degeneracy is essential in extracting the
information of the MH. Therefore, at $L <$ 30 km it is impossible to
resolve the MH while at $L \approx$ 60 km there is a range of energies where the
affect of MH could be, in principle, visible. At such a distance,
the `solar' suppression of the reactor $\bar{\nu}_e$ flux is near its maximum
and thus the higher frequency and lower amplitude  `atmospheric' oscillations become
more easily identified.

In order to explore the sensitivity of a potential measurement and simplify our
discussion, we assume a 40 GW thermal power of a reactor complex and
a 20 kT detector. In the absence of oscillations, the event rate per year at 1~km distance, $R$,
is estimated using the results of the Daya Bay experiment~\cite{dayabay} to be
$R = 2.5\times10^{8} /{\rm year}$.
At a baseline distance of $L$, the total number of events $N$ is then expected to be
$N = R \cdot T~(year) / L({\rm km})^2 \times \bar{P}(\bar{\nu_e}\rightarrow \bar{\nu_e})$,
where $\bar{P}(\bar{\nu_e}\rightarrow \bar{\nu_e})$ is the average
neutrino survival probability. Values of mixing angles and mass-squared
differences used in the simulation are taken from ~\cite{PDG,dayabay}:
\begin{eqnarray}
\sin^2 2\theta_{12} &=& 0.861^{+0.026}_{-0.022} \nonumber\\
\Delta m^2_{21} &=& (7.59\pm 0.21)\times 10^{-5} eV^2 \nonumber\\
\sin^2 2\theta_{23} &\sim& 1\nonumber\\
|\Delta m^2_{32}| &=& (2.43 \pm 0.13)\times 10^{-3} eV^2 \nonumber\\
\sin^2 2\theta_{13} &=& 0.092 \pm 0.017 ~(Daya~Bay)
\end{eqnarray}
For example, with  5 years running at 60 km, the total number of events
is about $10^5$. In addition, we assume $a=2.6$ in Eq.~\eqref{eq:res}. 
The reactor anti-neutrino spectrum was taken from Ref.~\cite{spectrum}. 
The fuel fractions of U$^{235}$, U$^{238}$, Pu$^{239}$, and Pu$^{241}$ 
are assumed to be 64\%, 8\%, 25\%, and 3\%, respectively.

\textcolor{black}{For the IBD measurement with such a detector, the majority of the backgrounds come from four types of events: the accidental coincidence events, the Li$^9$/He$^8$ 
decay events, the fast neutron events, and the geo-neutrino events. The accidental 
coincidence background can be determined from the experimental data with negligible
systematic uncertainties~\cite{dayabay_long,dayabay_cpc,dayabay_proceeding}. Both the Li$^9$/He$^8$ 
decay events and the fast neutron events are caused by cosmic muons. Such backgrounds are significantly suppressed in an experimental site situated deep underground, and their spectra are directly constrained by tagging the cosmic muons~\cite{dayabay_long,dayabay_cpc}. 
The geo-neutrino background with an energy spectrum of $E_{vis} <$ 2.5 MeV will 
give rise to about 3\% contamination extrapolated from the measured rate from 
KamLAND~\cite{KamLAND_geo} with a 40\% relative uncertainty.
Since geo-neutrinos originate 
from U and Th decays, their spectra are very well known and can be included 
into the spectrum analysis. 
Overall, we do not expect the backgrounds to pose a significant challenge
 in resolving the MH. While it will be important to include 
the effects of backgrounds in a sensitivity calculation for a realistic design, 
we did not include them in this study. } 

Fig.~\ref{fig:spectrum} shows the comparison of the IBD
energy spectrum (top panels) and the ratio of NH to IH spectrum (bottom panels)
w.r.t. $E_{vis} \approx  E_{\bar{\nu}} - 0.8 $ in MeV. It is important to note 
that we assumed a perfect absolute energy calibration and knowledge
of reactor IBD spectrum. 
Also, the \textcolor{black}{ideal spectrum without statistical fluctuations}
is considered in the left and middle panels. Compared with the case at known
$|\Delta m^2_{32}|$ with no uncertainty (left panels in Fig.~\ref{fig:spectrum}),
the difference between NH and IH
can be considerably reduced due to the lack of precise knowledge
of $|\Delta m^2_{32}|$ (middle panels in Fig.~\ref{fig:spectrum}). Furthermore, in
right panels of Fig.~\ref{fig:spectrum}, we show the realistic spectrum of NH
\textcolor{black}{with statistical fluctuations at 100 $kT\cdot year$ exposure together
with the ideal spectrum for the IH case.} The ratio of these two spectra is shown
in the bottom right panel.

In this section we have therefore identified the ambiguities associated with the uncertainty
of the  $|\Delta m^2_{32}|$ value in relation to the finite detector energy resolution. 
In particular, we have shown that,
under rather ideal conditions (perfect energy calibration, very long exposure, etc.),
the corresponding degeneracies can be overcome at intermediate distances ($\sim$ 60 km) and
in a limited range of energies.


\section{Extraction of the mass hierarchy}\label{sec:FT}

\begin{figure}[]
\centering
\includegraphics[width=90mm]{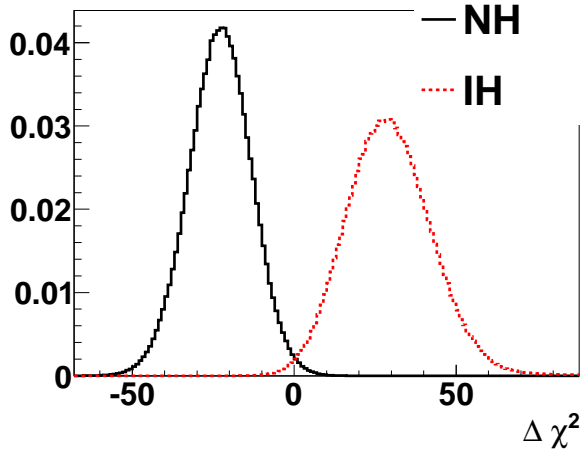}
\caption{The $\Delta \chi^2$ spectrum from Monte Carlo simulation. The NH (IH) represents the case when the nature is 
normal (inverted) hierarchy.}
\label{fig:chi2dis}
\end{figure}

In order to study the sensitivity of the mass hierarchy determination under these conditions,
we use the $\chi^2$ method together with Monte-Carlo simulations to compare the simulated
IBD energy spectrum of 100 $kT \cdot year$ exposure with the expected spectrum
in both NH and IH cases. The procedure is as follows.
First, the simulated spectrum was fit assuming NH by minimizing
\begin{equation}\label{eq:chi2}
\chi_{NH}^2 = \sum_i \frac{(S^i_{m} - S^i_{e~NH}(\Delta m^2))^2}{(\delta S^i_{m})^2} +\chi^2_{p} (\Delta m^2)
\end{equation}
with respect to $\Delta m^2$. Here, $S^i_m$ ($S^i_{e~NH}$) is the measured spectrum
(the expected spectrum with NH which depends on value of $\Delta m^2$) at the $i$th bin.
The $\delta S^i_m$ is the statistical uncertainty in the $i$th bin. The last term in Eq.~\eqref{eq:chi2}
is the penalty term from the most recent constrains of $|\Delta m^2_{32}|$ of MINOS
($|\Delta m^2| = 2.32 ^{+0.12}_{-0.08} \times 10^{-3} eV^2$~\cite{MINOS}). The value of
$\Delta m^2$ at the minimum $\chi^2$ is defined as $\Delta m^2_{min~NH}$. 
Second, the fit was repeat assuming IH to obtain $\chi_{IH}^2$ and $\Delta m^2_{min~IH}$.
Third, the difference in chi-square values ($\Delta \chi^2$) is defined as:
\begin{equation}
\Delta \chi^2 \equiv \chi_{NH}^2(\Delta m^2_{min~NH}) - \chi_{IH}^2(\Delta m^2_{min~IH}).
\end{equation}
\textcolor{black}{In this procedure, we have neglected the uncertainties of $\Delta m^2_{21}$, $\theta_{12}$,
and $\theta_{13}$, as we do not expect them to have a big impact on the MH resolution. 
First of all, we foresee that the precisions for these parameters will be significantly improved in the 
future. The uncertainty on $\theta_{13}$ will be determined by the final Daya Bay results to 
$\sim$5\%~\cite{dayabay_proceeding}. The precision of the $\theta_{12}$ and the $\Delta m^2_{21}$ can be 
improved in this medium-baseline measurement through the neutrino oscillation of 
solar term (last term in first line of Eq. 1). 
Moreover, the MH determination relies on the frequency measurement rather than 
the amplitude measurement of the neutrino oscillation. Therefore, it is less
sensitive to uncertainties of mixing angles. In addition, since the uncertainty of $\Delta m^2_{21}$ is much smaller than the 
changes in $\Delta m^2_{\phi}$, it will have negligible impact on the MH 
resolution as well. }

The distributions of $\Delta \chi^2$ for the true NH (black solid line) or IH (dotted red line) are
shown in Fig.~\ref{fig:chi2dis}. The area under each histogram is normalized to unity.
Furthermore, since the true value of $|\Delta m^2_{32}|$
is not known, the value of $|\Delta m^2_{32}|$ used in the simulated spectrum is randomly
generated according to the the most recent constrains of $|\Delta m^2_{32}|$ from MINOS.
Fourth, given a measurement with a particular value of $\Delta \chi^2$, the probability of
the MH being NH case can be calculated as $\frac{P_{NH}}{P_{NH}+P_{IH}}$. The $P_{NH}$ ($P_{IH}$)
is the probability density assuming the nature is NH (IH), which can be directly determined from
Fig.~\ref{fig:chi2dis}. Finally, the average probability can be calculated by evaluating 
the weighted average based on the $\Delta \chi^2$ distribution in Fig.~\ref{fig:chi2dis} 
assuming the truth is NH. A more detailed description on the average probability can be found in
Ref.~\cite{StatMH}. With 100 $kT \cdot year$ exposure with resolution parameter $a = 2.6$,
the average probability is determined to be 98.9\%.
Since this average probability is obtained by assuming a perfect knowledge of neutrino spectrum
as well as the energy scale, it represents the best estimate for the separation of mass hierarchy.

In order to relax the requirement of knowledge on energy scale and energy spectrum,
an attractive Fourier transform (FT) method was proposed recently
in Refs.~\cite{learned,zhan,zhan1}. In particular, in \cite{zhan} the quantity $(RL+PV)$ is introduced
\begin{equation}
RL = \frac{RV-LV}{RV+LV}  ~~~PV = \frac{P-V}{P+V} ~,
\end{equation}
where $P$ is the peak amplitude and $V$ is the amplitude of the valley in the
Fourier sine transform (FST) spectrum. There should be two peaks in the FST spectrum,
corresponding to $\Delta_{32}$ and $\Delta_{31}$, and the
labels $R,(L)$ refer to the right (left) peak.
Simulations in Ref.\cite{zhan1} show that the signs of $RL$ and $PV$ are related to the
hierarchy; positive for NH and negative for IH. In addition, in \cite{zhan1}
it was argued that value of $RL+PV$ is not sensitive to the detailed structure of the reactor
IBD spectrum nor to the absolute energy calibration.

\begin{figure}[]
\centering
\includegraphics[width=90mm]{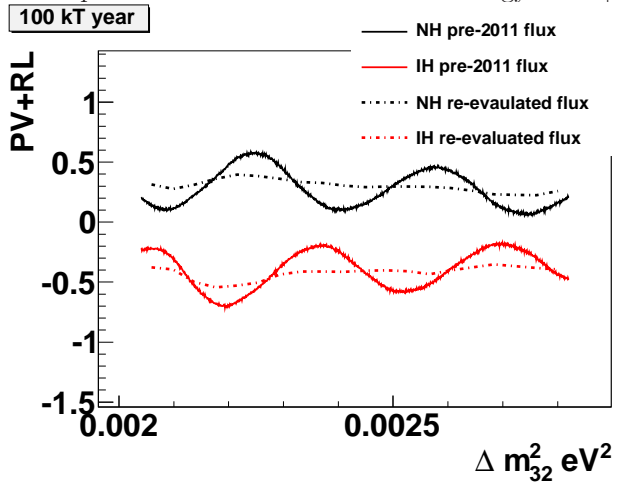}
\caption{ Values of $(RL+PV)$ for a range of $|\Delta m^2_{32}|$ and both
hierarchies are plotted for the 100 $kT\cdot year$ exposure with both pre-2011 flux
and the re-evaluated flux. }
\label{fig:FT}
\end{figure}

In Fig.~\ref{fig:FT}, we plot the central values of $(PV+RL)$ for a range of $|\Delta m^2_{32}|$
and for both hierarchies with the pre-2011 flux~\cite{spectrum,spec1,spec2,spec3,nspec3}
and the new re-evaluated flux~\cite{nspec2,nspec1,nspec3}. Although
the general feature of $(PV+RL)$ (positive for NH and negative for IH) is confirmed,
the $|\Delta m^2_{32}|$ dependence of $(PV+RL)$ value is shown to depend on
the choices of flux model. In addition, as we emphasized in
Fig.~\ref{fig:phase} when trying to determine the MH,
one should not use just one fixed value of $|\Delta m^2_{32}|$ for comparison
of the NH case with the IH case (as was done in Refs. \cite{zhan,zhan1})
but consider all possible values of  $|\Delta m^2_{32}|$ within
the current experimental uncertainties. The observed oscillation behavior with
pre-2011 flux would lead to a reduction in the probability to determine the MH.
With the Monte-Carlo simulation procedure using $(PV+RL)$, the average
probability is determined to be 93\% with the pre-2011 flux. Furthermore, the
average probability is expected to be smaller than that from
the full $\chi^2$ method in general, since the FT method utilizes less information
(e.g. only heights of peaks and valleys) in order to reduce the requirement
in energy scale determination.  Fig.~\ref{fig:FT} shows that
a good knowledge of the neutrino flux spectrum is desired to
correctly evaluate the probability of MH determination with the FT method.

\section{Challenges of the energy scale} \label{sec:e-scale}


As stressed in the discussion of Fig.~\ref{fig:phase}, in the energy
interval $E_{vis} = 2-4$ MeV (at $L$ = 60 km) the quantity  $\Delta m^2_{\phi}$ changes
significantly with respect to the uncertainty in $|\Delta m_{32}|^2$. The lower limit of that region is caused by
the smearing of the fast oscillations of the observed spectrum
due to the finite detector energy resolution,
while the upper limit is caused by the degeneracy, i.e. by the fact that
 $\Delta m^2_{\phi}$ becomes almost independent of energy from that value on.
All of these are then reflected in the FT analysis.
Although the FT method does not require an absolute calibration
of energy scale~\cite{zhan1}, a precision calibration of the relative 
energy scale is extremely important. 
A small non-linearity of the energy scale characterization 
can lead to a substantial reduction of the discovery potential.

To illustrate this point, we consider the case corresponding
to IH, and assume that (due to imperfect understanding of the detector performance) the reconstructed energy $E_{rec}$ is
related to the real energy $E_{real}$ by the relation
\begin{equation}
E_{rec} = \frac{2|\Delta' m^2_{32}| + \Delta m^2_{\phi}(E_{\bar{\nu}},L)}{2|\Delta m^2_{32}|  - \Delta m^2_{\phi}(E_{\bar{\nu}},L)}E_{real}~.
\label{eq:escale}
\end{equation}
(Here we use the notation $|\Delta' m^2_{32}|$ and $|\Delta m^2_{32}|$ to emphasize
the fact that  $|\Delta m^2_{32}|$ is known only within a certain error.)
If the energy scale is distorted according to this relation, and that distortion is not
included in the way the reconstructed energy is derived from the data,
the pattern of the disappearance probability regarding the atmospheric term 
will be exactly the same as in the NH case.
This can be seen as:
\begin{widetext}
\begin{equation}
\cos\left( (2|\Delta m^2_{32}| - \Delta m^2_{\phi}(E_{\bar{\nu}},L)) \frac{L}{E_{real}}\right) 
= \cos\left( (2|\Delta' m^2_{32}| + \Delta m^2_{\phi}(E_{\bar{\nu}},L)) \frac{L}{E_{rec}}\right)
\end{equation}
\end{widetext}
from Eq.~\eqref{eq:osci}.
In this case the analysis of the spectrum would lead
to an obviously wrong MH. Since the exact value of  $|\Delta m^2_{32}|$ is not
known, we must consider in Eq.~\eqref{eq:escale} all allowed values of  $|\Delta' m^2_{32}|$
including those that minimize the ratio $E_{rec}/E_{real}$.

\begin{figure}[]
\centering
\includegraphics[width=90mm]{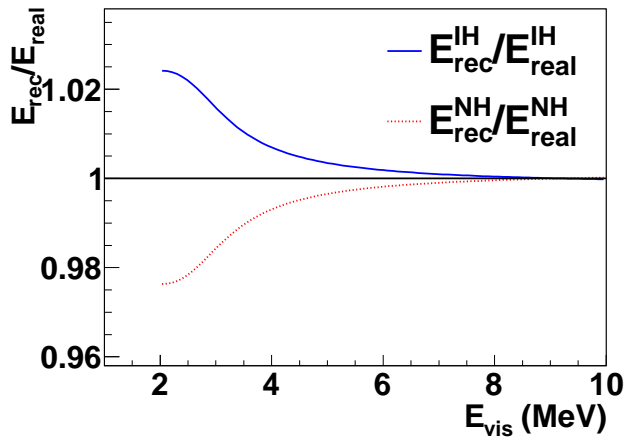}
\caption{The ratio of $E_{rec}$ to $E_{real}$ for the case of IH based
on Eq.~\eqref{eq:escale} (solid line) is shown w.r.t the visible energy $E_{vis}$.
The dotted line shows the ratio of $E_{rec}$ to $E_{real}$ for the case of NH. }
\label{fig:energy}
\end{figure}

Fig.~\ref{fig:energy} shows the ratio $E_{rec}/E_{real}$ versus the visible
energy (solid line) with the energy scale distortion described by Eq.~\eqref{eq:escale}
where  $|\Delta' m^2_{32}|$ was chosen so that this ratio is one 
at high $E_{vis}$.
Comparing the medium energy region (2 MeV $<E_{vis}<$ 4 MeV) with
the higher energy region ($E_{vis} >$ 4 MeV), the
average $E_{rec} /E_{real}$ is larger than unity by only about 1\%. In addition, the same
argument similar to Eq.~\eqref{eq:escale} applies to the NH case as well. The ratio
$E_{rec}/E_{real}$ versus the visible energy (dotted line) of NH is also shown
in Fig.~\ref{fig:energy}. Therefore, to ensure
the MH's discovery potential from such an experiment, the non-linearity of energy scale ($E_{rec}/E_{real}$)
needs to be controlled to a fraction of 1\% in a wide range of $E_{vis}$. This requirement should
be compared with the current state-of-art 1.9\% energy scale uncertainty from
KamLAND~\cite{kamland1}. Therefore, nearly an order of magnitude improvement in the
energy scale determination is required for such a measurement to succeed.


\section{Uncertainties in $|\Delta m^2_{32}|$}

\begin{figure}[]
\centering
\includegraphics[width=90mm]{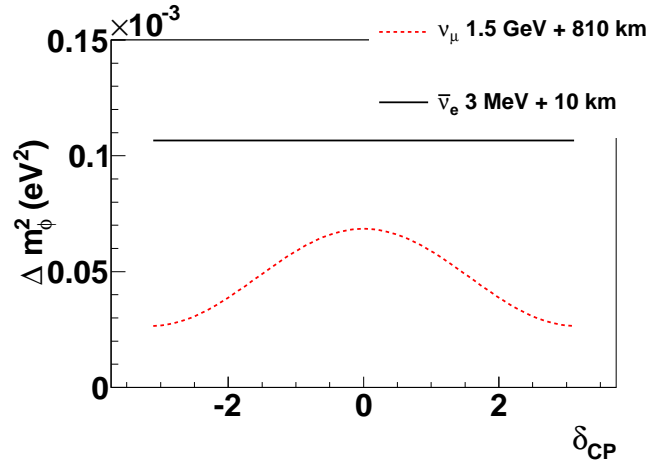}
\caption{The dependence of effective mass-squared difference $\Delta m^2_{ee\phi}$ (solid line)
and $\Delta m^2_{\mu\mu\phi}$ (dotted line) w.r.t. the value of $\delta_{CP}$ for $\bar{\nu}_{e}$
and $\nu_{\mu}$ disappearance measurements, respectively.}
\label{fig:dyna}
\end{figure}

The current primary method to constrain $|\Delta m^2_{32}|$ is the $\nu_{\mu}$ disappearance experiment.
However, similar to the $\bar{\nu}_e$ disappearance case as in Eq.~\ref{eq:osci}, the $\nu_{\mu}$ disappearance measurement
in vacuum~\footnote{In practice, the uncertainty in the matter effect would introduce only a systematic
uncertainty. The strength of the effect in $\nu_\mu$ disappearance is close to that of
changing $|\Delta m^2_{32}|$ by a few times of $10^{-6} eV^2$.}
would also measure an effective mass-squared difference rather than $|\Delta m^2_{32}|$ directly.
The corresponding effective mass-squared difference is smaller than that in the $\bar{\nu}_e$ case,
basically since in the Eq.~\eqref{eq:phi} the cosine squared of $\theta_{12}$ is replaced by
the sine squared. Also, in this case, the effective mass-squared difference
 will depend not only on $\Delta_{21}$, $\theta_{12}$, but also on
$\theta_{13}$, $\theta_{23}$, as well as on the unknown CP violation phase $\delta_{CP}$.
The effective mass-squared differences from $\nu_{\mu}$ and $\nu_e$  disappearance
w.r.t. the value of $\delta_{CP}$ are shown in Fig.~\ref{fig:dyna}. The difference in $\Delta m^2_{\phi}$
between the $\nu_{\mu}$ and $\nu_e$ channels actually opens a new path to determine the MH. This
possibility was discussed earlier in Refs. \cite{Mina, Nuno}. It was stressed
there that the difference in frequency shifts $2\Delta_{32} \pm \phi$ has opposite
signs for the $\bar{\nu}_e$ and $\nu_{\mu}$ disappearance in the normal or inverted
hierarchies. Such a measurement would require that $2\Delta_{32} \pm \phi$ is measured
to a fraction of $\Delta m^2_{ee\phi}-\Delta m^2_{\mu\mu\phi}$ level ($5\times10^{-5}~eV^2$) in both channels.
In the current $\sim$ 60 km configuration, the knowledge of $|\Delta m^2_{32}|$ enters through
the penalty term in Eq.~\eqref{eq:chi2}. Therefore, in order for knowledge of $|\Delta m^2_{32}|$
to have a significant impact to the determination of MH, the $\Delta_{32} \pm \phi$ in $\nu_{\mu}$ channel
should also be measured to a fraction of $\Delta m^2_{ee\phi}-\Delta m^2_{\mu\mu\phi}$ level,
which is well beyond the reach of T2K~\cite{t2k} and NO$\nu$A~\cite{nova} $\nu_{\mu}$ 
disappearance measurements~\footnote{\textcolor{black}{The projected 1-$\sigma$ uncertainties on $|\Delta m^2|=|\Delta m^2_{32}\pm\Delta m^2_{\mu\mu\phi}/2|$
from T2K and NO$\nu$A are about $5.3\times 10^{-5}$ eV$^2$.}}.


\section{Conclusions}

In summary, the sensitivity of determining the neutrino mass hierarchy
using the reactor neutrino experiment at $\sim$60 km is explored and its
challenges are discussed. Such a measurement is difficult due to the finite
detector energy resolution, to the necessity of the accurate
absolute energy scale calibration, and to  degeneracies related to the current experimental
uncertainty of $|\Delta m_{32}^2|$. The key to the success of such a measurement
is to control the systematic uncertainties. We show here that one
must understand the non-linearity of the detector energy scale to a few tenths of percent,
which requires nearly an order of magnitude of improvement in
the energy scale compared to the current state-of-art limit, 1.9\% from KamLAND.


\section{Acknowledgments}
We would like to thank Liang Zhan and Jiajie Ling for fruitful discussions.
This work was supported in part by Caltech, the National Science
Foundation, and the Department of Energy  under contracts DE-AC05-06OR23177, under which 
Jefferson Science Associates, LLC, operates the Thomas Jefferson
National Accelerator Facility, and DE-AC02-98CH10886.

\bibliographystyle{unsrt}
\bibliography{IBD_MH}{}

\end{document}